\documentclass[showpacs, pra,onecolumn,preprintnumbers ,amsmath, amssymb, superscriptaddress, aps]{revtex4-2}

\usepackage{color}
\usepackage{amsmath,amssymb}
\usepackage{pifont}
\usepackage{amssymb}  
\usepackage{bbold}
\usepackage{float}
\usepackage{subfloat}
\usepackage[caption=false]{subfig}
\usepackage{tikz}
\usepackage{makecell}
\usepackage{subfig}
\usepackage{pifont}   
\usepackage{graphicx} 
\graphicspath{{Figures2/}}
\usepackage{dcolumn}  
\usepackage{bm}       
\usepackage{multirow} 
\usepackage{placeins}
\usepackage[colorlinks]{hyperref}
\usepackage{mathtools}
\usepackage{appendix}

\def \be{\begin{align}}
	\def \ee{\end{align}}
\def \bea{\begin{eqnarray}}
	\def \eea{\end{eqnarray}}

\begin{document}
	
	\title{Laser-assisted tunneling and Hartman effect in graphene under scalar potential and exchange fields
	}
	
	\author{Rachid El Aitouni}
	\affiliation{Laboratory of Theoretical Physics, Faculty of Sciences, Choua\"ib Doukkali University, PO Box 20, 24000 El Jadida, Morocco}
	\author{Ahmed Jellal}
	\email{a.jellal@ucd.ac.ma}
	\affiliation{Laboratory of Theoretical Physics, Faculty of Sciences, Choua\"ib Doukkali University, PO Box 20, 24000 El Jadida, Morocco}

\author{Pablo Díaz}
\affiliation{Departamento de Ciencias F\'{i}sicas, Universidad de La Frontera, Casilla 54-D, Temuco 4811230, Chile}  
\author{David Laroze}
\affiliation{Instituto de Alta Investigación, Universidad de Tarapacá, Casilla 7D, Arica, Chile}

	\begin{abstract}
		
	{We study the tunneling effect of Dirac fermions in a graphene sheet by introducing a potential barrier in a region of width $D$ exposed to laser field. This sheet is placed on a boron nitride/ferromagnetic substrate such as cobalt or nickel. By using the Floquet theory, we determine the solutions of the energy spectrum. We calculate the transmission and reflection coefficients by applying the boundary conditions along with the transfer matrix method. These coefficients help determine their probabilities by current densities and group delay times by their phases.
	We numerically show that the laser field plays a crucial role in this structure, as it completely suppresses Klein tunneling compared to the case without laser. Furthermore, in contrast to the Hartman effect, the group delay time becomes dependent on the barrier width with the appearance of additional peaks. This suggests that fermion-field interactions cause additional delays within the barrier and also help to reduce spin coupling. Adding BN layers increases the interval of transmission suppression and completely eliminates coupling after the addition of three BN layers. Total reflection is observed for incident fermions with an angle less than $-1$ or greater than one.
	}

	\end{abstract}

	\maketitle

\section{Introduction}
Graphene is a two-dimensional carbon-based material \cite{Novos2004}. It is known to be both rigid and flexible \cite{prop}, and it absorbs only \(2.3\%\) of light \cite{absor}. The electrons in graphene move at a speed 300 times slower than the speed of light \cite{mobil,mobil2}, which is why they are called massless Dirac fermions \cite{masless}. These unique properties make graphene a promising material for use in electronic devices. However, a major challenge is the lack of a band gap between its valence and conduction bands \cite{zero,zero1}. These bands meet at six points, called Dirac points, resulting in no energy gap.
To address this issue, several methods have been proposed to create a bandgap in graphene. One approach is to deform the graphene structure \cite{def1,def2}. Another method is doping, where other elements are added to graphene \cite{dopage}. Applying external fields, such as magnetic fields \cite{mag1,mag3,mag4,magneticfield} or lasers \cite{bis2}, is also a common strategy. Additionally, depositing graphene on specific substrates \cite{substrat} has shown potential in creating a bandgap. These methods aim to modify the electronic properties of graphene,  making  more suitable for electronic applications.
Even after solving the bandgap problem, another challenge arises—Klein tunneling \cite{klien1,klien2,klienexp}. This phenomenon allows fermions to pass through potential barriers even when their energy is less than the barrier height. This behavior is a major obstacle to the use of graphene in electronic devices because it prevents the effective control of electron flow.Overcoming these two challenges—bandgap creation and Klein tunneling—is essential for the practical use of graphene in electronics. By solving these problems, researchers can unlock the full potential of graphene. This could lead to the development of faster transistors, flexible displays, and highly efficient sensors. The  unique properties of graphene, such as high electron mobility, mechanical strength, and optical transparency, make it a promising candidate for next-generation technologies. Overcoming these challenges will pave the way for innovative applications and advances in electronics.


Time-oscillating barriers \cite{oscil1,oscil2,timepot,timepot2,doubletemps} create a quantized energy spectrum, where the barrier displays multiple energy levels. This setup allows for photon exchange with fermions as they pass through the barrier, resulting in two types of transmission processes: one without photon exchange and another with photon exchange. Laser irradiation can also produce these two processes. Studies show that increasing the laser field intensity reduces transmission and effectively suppresses Klein tunneling \cite{Elaitouni2023,Elaitouni2023A,ELAITOUNI2024,doublelaser}.
The time it takes for fermions to cross the barrier, called the group delay time, is a key factor in assessing the practicality of this technique. In 1962, Hartman studied this delay using a simple barrier \cite{Hartman1962}. He discovered that beyond a certain barrier width, the traversal time no longer depends on the barrier thickness. This phenomenon is now known as the Hartman effect. The group delay is calculated using the phase of the transmission and reflection coefficients. Measuring this delay accurately is important for applications in telecommunications and the development of more efficient electronic components.
Understanding these effects allows the design of improved systems to control electron flow and enhance device performance. By studying time-oscillating barriers and laser irradiation, we can optimize transmission and delay time. This knowledge is critical to developing advanced technologies with greater precision and efficiency, enabling the creation of faster, more reliable electronic devices and the optimization of communication systems.


We study the tunneling behavior in graphene subjected to a potential barrier, a laser field within a region of width $D$, and the exchange field induced by a ferromagnetic substrate. The graphene sheet is deposited on a substrate, either nickel or cobalt, and separated by boron nitride (BN) layers, with configurations consisting of one (1), two (2), or three (3) BN layers. 
We study the combined influence of the bandgap induced by the potential barrier and the exchange interaction with the substrate, as well as the effect of the laser field on both Klein tunneling and the Hartman effect. We show that laser irradiation can be used to reduce transmission and suppress Klein tunneling compared to the case without the laser field \cite{Tepper2021}. It also reduces the effect of spin coupling. These results highlight the importance of laser field modulation as a control of the electron behavior in graphene-based devices.
It is found that transmission is imperfect even when the incident energy exceeds the barrier height, unlike in the case of a simple barrier without a substrate \cite{Klein1929}. This imperfection is due to the interaction between the fermions and the laser field. The interval at which transmission is canceled is also affected by the number of boron nitride (BN) layers, with the interval increasing as more layers are added, with a corresponding decrease in transmission. The group delay time, which depends on the barrier width, indicates the absence of the Hartman effect \cite{Hartman1962}. Under laser irradiation, the group delay time decreases for normal incidence but increases for other angles of incidence. The appearance of new peaks, comparable to the scenario without laser irradiation \cite{Tepper2021}, suggests an additional delay for incident fermions at certain angles, further highlighting the role of the laser field in influencing the transmission dynamics.

{In the present study, we aim to provide a theoretical framework for understanding how a single barrier and a laser field affect tunneling in graphene. While our model provides valuable insights, we recognize that real-world applications will need to address practical challenges. On the other hand, graphene can be created by chemical vapor deposition (CVD) and deposited on a layer of hexagonal boron nitride (h-BN) \cite{Li2009}. A ferromagnetic material such as cobalt (Co) or nickel (Ni) can then be sputtered onto the h-BN substrate \cite{Dean2010}. In addition, a tunable potential barrier can be created by patterning metallic gates on the graphene layer and applying a voltage \cite{Huard2007}. The system can then be exposed to a laser field to tune the electronic properties of graphene \cite{Oka2009}. Perhaps tunneling effects and group delay times can be analyzed using techniques such as scanning tunneling microscopy (STM), Raman spectroscopy, and transport measurements \cite{Novos2004}.

}

The present paper is organized as follows. In Sec. \ref{TM}, we introduce a theoretical model and determine the energy spectrum using the Floquet approximation. In Sec. \ref{TR}, we use boundary conditions and the transfer matrix approach along with current densities to determine the transmission and reflection probabilities. These are used to derive the corresponding group delay times in Sec. \ref{GDT}. Sec. \ref{NR} presents our numerical analysis based on different conditions of the physical parameters.  Sec. \ref{Con} gives a summary and conclusion of our results.
 
\section{THEORETICAL MODEl}\label{TM}

In the tight-binding model approximation \cite{Wallace1947}, the Hamiltonian describing the motion of an electron in graphene is given by
\begin{align}
	\mathcal{H}_{0}= v_F (\tau  {\sigma_x} {p_x} + {\sigma_y}{p_y})
\end{align}
where $p_x$ and $p_y$ are the momentum vector components, $\sigma_i$($i=x,y,z$) are the Pauli matrices, $\tau=1$ for the ${K}$ valley and $\tau=-1$ for the ${K'}$ valley, $v_F$ is the Fermi velocity.
After depositing graphene on a ferromagnetic substrate of nickel or cobalt, either directly or separated by a few layers of boron nitride (BN), the Hamiltonian becomes spin-dependent due to spin-orbit coupling, with the appearance of a gap caused by interaction with the substrate. Therefore, the modified Hamiltonian takes the following form
\begin{align}\label{Ht}
	\mathcal{H}=\mathcal{H}_{0}+\mathcal{H}_{\Delta}+\mathcal{H}_{so} \end{align}
and new terms have the forms 
	\begin{align}
	\mathcal{H}_{\Delta}= \Delta{\sigma_z}, \quad \mathcal{H}_{so}=\frac{\tau}{2}\left[\lambda_A({\sigma_z+\sigma_0})+\lambda_B({\sigma_z-\sigma_0}) \right]
	\end{align}
	such that $\lambda_A$ and $\lambda_B$ denote the spin-orbit coupling parameters for the two sublattices, and ${\sigma_0}$ is the $2 \times 2$ identity matrix in the pseudospin space. The band gap width is then $\tau \beta(\lambda_B + \lambda_A) + 2\Delta$, where $\beta = 1(-1)$ for spin up (down).
	
	\begin{figure}[ht]
		\centering
		\includegraphics[scale=0.2]{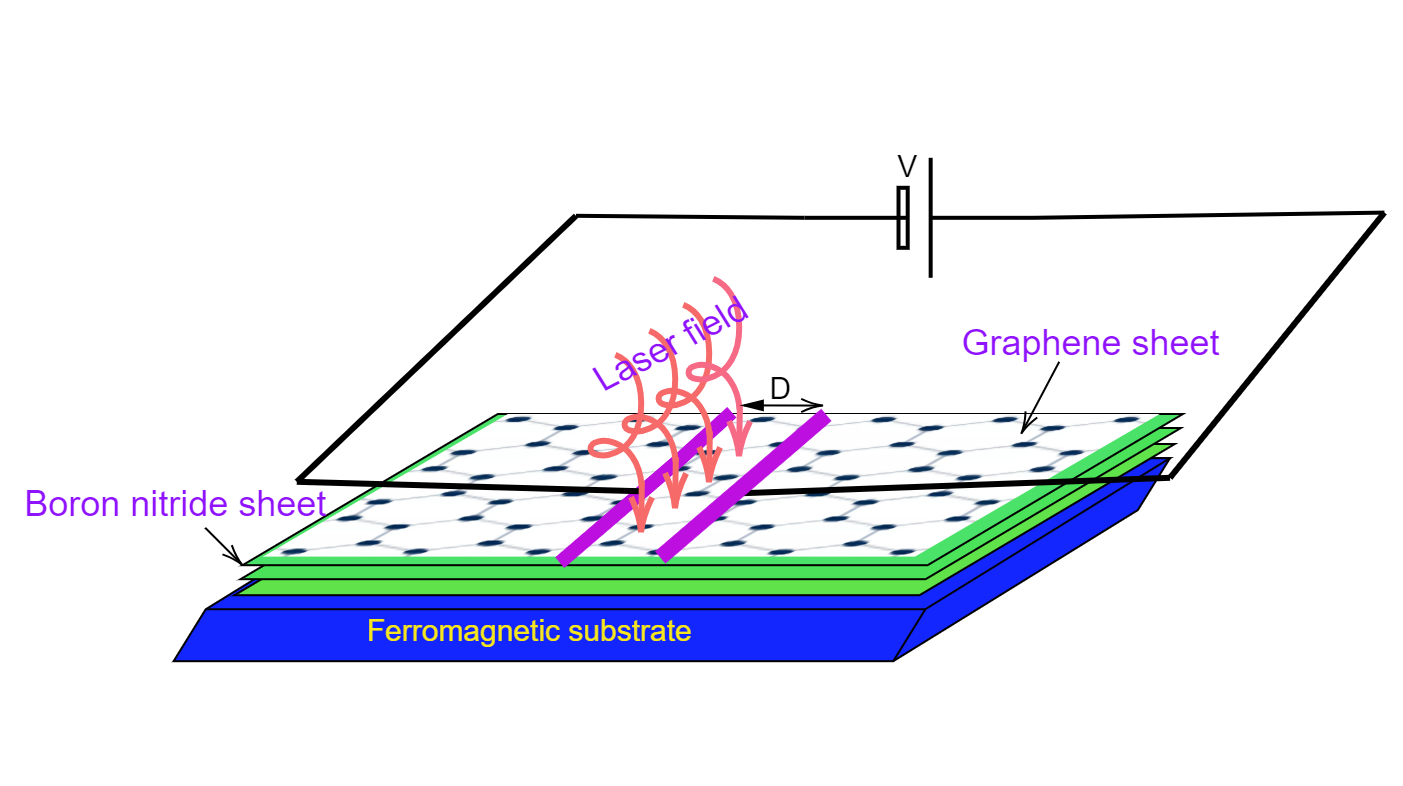}
		\caption{Schematic of a graphene sheet deposited on a ferromagnetic substrate, separated by layers of boron nitride (BN). A region of width $D$ under the influence of a potential barrier and also irradiated by a monochromatic laser field.} \label{str}
	\end{figure}
	
	We consider a layer of graphene deposited on a ferromagnetic substrate of cobalt or nickel, separated by layers of boron nitride (BN) as depicted in see Fig. \ref{str}. In a region of width $D$, we apply a potential barrier of height $V$ that is irradiated by a laser field generated by a sinusoidal electric field $E=F \sin(\omega t)$ of frequency $\omega$ and intensity $F$. From the relation $E=-\frac{\partial A}{\partial t}$, we get the associated vector potential \cite{dipole}
\begin{align} A_L(t)=
	\begin{cases}
		A_0 \cos(\omega t) &0<x<D\\
		0,& \text{otherwise}
	\end{cases}
\end{align} 
where   $A_0=\frac{F}{\omega}$ is the  amplitude. We also consider the scalar potential defined by
\begin{align}
V(x)=
	\begin{cases}
		V &0<x<D\\
		0,& \text{otherwise}
	\end{cases}
\end{align}
Combining all parts, we write the Hamiltonian that describes the motion of an electron through the structure engineered in Fig. \ref{str} as
\begin{equation}
		\mathcal{H}= \begin{pmatrix}
		\beta\lambda_A+\Delta+V &  v_F (\tau p_x -ip_y-i{e}A_L) \\
		v_F (\tau p_x +ip_y+i{e}A_L)&-\beta\lambda_B-\Delta+V\\
	\end{pmatrix} 
	\label{hprox}
\end{equation}	
As the laser field oscillates over time, the energy spectrum is quantized by the photon exchange between the laser field and the fermions. Consequently, the wave function in the three regions is expressed as \cite{oscil1,oscil2,timepot,timepot2,doubletemps}
\begin{equation}
	\psi_{j}(x,y,t)=\phi_j(x,y,0)\sum_{m=-\infty}^{\infty}J_m(\alpha)e^{-i(E+m\omega \hbar) t/\hbar}
\end{equation}
and we set the parameter $\alpha=\frac{e F v_F}{\hbar\omega^2}$.  To determine \(\phi_j(x, y, 0)\) we must treat each region separately. This approach allows us to account for the different physical conditions and boundary conditions in each region, ensuring that the wave function solutions accurately reflect the behavior of the electron in different parts of the system. By solving for \(\phi_j(x, y, 0)\) region by region, we can then adjust the solutions at the boundaries to obtain a complete and consistent description of the system.

The applied barrier divides the system into three distinct regions. In regions 1 and 3 we have pristine graphene, while in the intermediate region graphene is affected by both a laser field and a scalar barrier. To determine the wave functions in all three regions, we solve the eigenvalue equations corresponding to each region, taking into account the spin-orbit coupling and the effects of the substrate. The eigenspinors in regions 1 and 3 can be obtained as  \cite{Elaitouni2022,Elaitouni2023}
\begin{align}
	&\psi_{1}(x,y,t)=e^{i k_{y} y} \sum_{m,l=-\infty}^{\infty}\left[
	\begin{pmatrix}
		1\\	\xi_l
	\end{pmatrix}
	e^{i k_lx}\delta_{m,0}+r_l 
	\begin{pmatrix}
		1\\-\xi^*_l
	\end{pmatrix}
	e^{-i k_l x}\right]\delta_{m,l}e^{-i(E+m\hbar\varpi) t/\hbar}\\
	&\psi_{3}(x,y,t)= e^{i k_{y} y}\sum_{m,l=-\infty}^{\infty}t_l 
	\begin{pmatrix}
		1\\ \xi_l
	\end{pmatrix}
	e^{i k_l x}\delta_{m,l}e^{-i(E+m\hbar\varpi) t/\hbar}
\end{align}
and the complex has the form 
\begin{align}
	\xi_l=	\frac{\hbar v_F (\tau k_l - 
		i k_y)}{\beta \lambda_B + \Delta + 
		E+ l \hbar \omega}.
\end{align}
The associated energies are given by
\begin{equation}
	E+m\hbar\omega=\lambda_{ex}^- \pm s_1^m\sqrt{\hbar^2 v_F^2(k_l^2+k_y^2)+\left(\lambda_{ex}^+ +\Delta \right)^2}
	\label{eprox1}
\end{equation}
where $k_l$ is the wave vector along the propagating direction $x$, 
 $s_1^m=\text{sign}(E+m\hbar\omega-\lambda_{ex})$ and $\lambda_{ex}^\pm=\beta(\lambda_A\pm\lambda_B)/2$. Here, the band gap edges are defined by $(\beta\lambda_A+\Delta)$ and $-(\beta\lambda_B+\Delta)$.

In region 2, where the scalar barrier is exposed to laser irradiation, the eigenvalue equation allows us to obtain the following coupled equations
\begin{align}
	\hbar v_F \left[-i\partial_x-i\left(k_y+m\frac{\omega}{v_F}\right)\right]\phi_{2j}(x,y)&=\left(E+m\hbar \omega-\Delta^A\right)\phi_{1j}(x,y)\\
	\hbar v_F \left[-i\partial_x+i\left(k_y+m\frac{\omega}{v_F}\right)\right]\phi_{1j}(x,y)&=\left(E+m\hbar \omega+\Delta^B\right)\phi_{2j}(x,y)
\end{align}
where we have defined $\Delta^A=\Delta+\beta \lambda_A+V$ and $\Delta^B=\Delta+\beta \lambda_B-V$. These equations can be manipulated to arrive at the solution
\begin{align}
	\psi_2(x,y,t)=\sum_{m,l=\infty}^{\infty}\left[a_{l}\begin{pmatrix}
		1\\
		\zeta_l
	\end{pmatrix}e^{iq_lx}+b_{l}\begin{pmatrix}
		1\\
		-\zeta^*_l
	\end{pmatrix}e^{-iq_l x}\right]e^{ik_yy}J_{m-l}(\alpha)e^{-i(E+m\hbar\omega)t/\hbar}
\end{align}
with the complex number 
\begin{align}
	\zeta_l=\frac{\hbar v_F \left[\tau q_l - 
		i \left(k_y+l\frac{\omega}{v_F}\right)\right]}{\beta \lambda_B + \Delta +
		E+ l \hbar \omega- V}
\end{align}
and $\beta=1(-1)$ stands for spin up (down).
The corresponding electronic band structure is then given by the dispersion relation
\begin{equation}
	E+l\hbar\omega=\lambda_{ex}^- \pm s_2^m\sqrt{\hbar^2 v_F^2\left[q_l^2+\left(k_y+l\frac{\omega}{v_F}\right)^2\right]+\left(\lambda_{ex}^+ +\Delta \right)^2}
	\label{eprox}
\end{equation}
where  $s_2^m=\text{sign}(E+m\hbar\omega-V-\lambda_{ex})$.
In the following sections, we will explore how the energy spectrum solutions can be used to analyze the tunneling effect in the current system. These will help to gain a deeper insight into the behavior of electrons as they cross potential barriers, shedding light on the mechanisms underlying quantum tunneling. More precisely, we will show that the interplay between the system parameters and the tunneling will provide a comprehensive understanding of the electron transport in the given setup.

\section{Transmissions}\label{TR}


To compute the transmission and reflection coefficients, we impose continuity of the wave functions at the two interfaces of the barrier, ensuring that \(\psi_1(x=0,y,t) = \psi_2(x=0,y,t)\) and \(\psi_2(x=D,y,t) = \psi_3(x=D,y,t)\).  From these continuity conditions, we derive four equations that govern the behavior of our system
\begin{align}
	&\delta_{m,0}+r_m=\sum_{l=-\infty}^{+\infty}(a_l+b_l)J_{m-l}(\alpha)\\
		&\delta_{m,0}\xi_m-r_m\xi^*=\sum_{l=-\infty}^{+\infty}(a_l\zeta_l-b_l\zeta^*_l)J_{m-l}(\alpha)\\
	&\sum_{l=-\infty}^{+\infty}(a_le^{iq^l_x D}+b_le^{-iq^l_x D})J_{m-l}(\alpha)=t_m\delta_{m,l}\\
		&\sum_{l=-\infty}^{+\infty}(a_l\zeta_le^{iq^l_x D}-b_l\zeta^*_le^{-iq^l_x D})J_{m-l}(\alpha)=t_m\xi_m\delta_{m,l}.
\end{align}
Each equation contains an infinite number of modes. This shows the complex nature of the wave functions when a potential barrier is exposed to a laser field. The complexity includes potential scattering and interference effects. By solving these equations, we can find the transmission and reflection coefficients. These coefficients describe how the incident electron waves split into transmitted and reflected parts as they interact with the barrier.
To facilitate the solution of the system, we use the transfer matrix method. This approach allows us to express the above set in the following way
\begin{align}\label{Tmat}
	\binom{\delta_{m,0}}{r_m}
	=\mathbb{M}\begin{pmatrix}
		t_m\\
		\mathbb{0}_m
	\end{pmatrix}
\end{align}
where  $\mathbb{0}_m$ being a zero vector, and $\mathbb{M}$ is  the transfer matrix linking region 1 to region 3
\begin{align}\label{M123}
	\mathbb{M}&=\mathbb{M}(1,2)\cdot\mathbb{M}(2,3)
	=\begin{pmatrix}
		\mathbb{M}_{11}&	\mathbb{M}_{12}\\
		\mathbb{M}_{21}&	\mathbb{M}_{22}
	\end{pmatrix}
\end{align}
involving the matrices $\mathbb{M}(1,2)$ and $\mathbb{M}(2,3)$, which link regions (1,2) and (2,3), respectively, 
	\begin{align}
	&\mathbb{M}(1,2)=\begin{pmatrix}
		\mathbb{I}&\mathbb{I}\\
		\mathbb{N}^+&\mathbb{N}^-
	\end{pmatrix}^{-1}\cdot \begin{pmatrix}
		\mathbb{J}&\mathbb{J}\\
		\mathbb{G}^+&\mathbb{G}^-
	\end{pmatrix}\\
	&\mathbb{M}(2,3)=\begin{pmatrix}
		\mathbb{Q}&\mathbb{Q}\\
		\mathbb{R}^+&\mathbb{R}^-
	\end{pmatrix}^{-1}\cdot \begin{pmatrix}
		\mathbb{I}&\mathbb{I}\\
		\mathbb{N}^+&\mathbb{N}^-
	\end{pmatrix}
\end{align}
and their  elements are given by
\begin{align}
	&(\mathbb{N}^\pm)_{m,l}=\pm\xi_m^{\pm 1}\delta_{m,l}\\
	&(\mathbb{J})_{m,l}=J_{m-l}(\alpha)\\
	&(\mathbb{G}^\pm)_{m,l}=\pm \zeta_m^{\pm 1} J_{m-l}(\alpha)\\
	&{(\mathbb{Q}^\pm)_{m,l}=e^{\pm iq_lD}J_{m-l}(\alpha)\delta_{m,l}}\\
	&{(\mathbb{R}^\pm)_{m,l}=\pm\zeta_m^{\pm 1}e^{\pm iq_lD}J_{m-l}(\alpha)\delta_{m,l}}
\end{align}
Using the expressions provided in \eqref{Tmat}, the transmission and reflection coefficients for different modes can be calculated as functions of the system parameters
\begin{align}
	&t_m=\mathbb{M}_{1,1}^{-1}\cdot\delta_{m,0}\\
	&	r_m=\mathbb{M}_{2,1}\cdot t_m=\mathbb{M}_{2,1}\cdot\mathbb{M}_{1,1}^{-1}\cdot\delta_{m,0}.
\end{align}


To deal with the infinite size of the transfer matrix $\mathbb{M}$, we approximate it by truncating the series and considering a finite number of terms from $-N$ to $N$. Here, $N$ is chosen to satisfy the condition $N > \alpha$, as mentioned in \cite{timepot,timepot2}. This approximation allows the series to be expressed in the following form
\begin{align}
	t_{-N+k} = \mathbb{M}^{-1}_{11}[k+1, N+1], \quad k = 0, 1, \cdots, 2N.
\end{align}
In this representation, $t_{-N+k}$ corresponds to certain elements of the inverse matrix $\mathbb{M}^{-1}_{11}$, where the indices $[k+1, N+1]$ are determined by the chosen truncation range. By limiting the series to $2N+1$ terms, we ensure computational tractability while maintaining sufficient accuracy for practical calculations. The choice of $N$ is crucial, as it directly affects the balance between accuracy and computational efficiency in the analysis.

To calculate the transmission and reflection coefficients, we use the incident, transmitted, and reflected current densities. This is accomplished by applying the continuity equation
\begin{align}
	\nabla \cdot \vec{J} + \frac{\partial |\psi(x,y,t)|^2}{\partial t} = 0
\end{align}
which allows us to determine the  current densities
\begin{align}
&	J^\text{inc}_0=v_F(\xi_0+\xi^*_0)\\
&	J^\text{tra}_l=v_Ft^*_lt_l(\xi_l+\xi^*_l)\\
&	J^\text{ref}_l=v_Fr^*_lr_l(\xi_l+\xi^*_l).
\end{align}
These can be used to derive  the transmission and reflection probabilities as follows
\begin{align}
&	T_l=\frac{|J^\text{tra}_l|}{|J^\text{inc}_0|}=|t_l|^2\\
&	R_l=\frac{|J^\text{ref}_l|}{|J^\text{inc}_0|}=|r_l|^2.
\end{align}
and it is clear that the total transmission and reflection are obtained by summing the contributions of all individual modes. They are given by
\begin{align}
&T=\sum_{l}T_l,\quad R=\sum_{l}R_l
\end{align}	
Given the challenges associated with analyzing all possible transmission modes, we will focus on three specific modes that correspond to the first three bands. These modes are defined as 
\begin{align}    t_{-1} = \mathbb{M}_{1,1}^{-1}[1,3], \quad t_{0} = \mathbb{M}_{1,1}^{-1}[2,3], \quad t_{1} = \mathbb{M}_{1,1}^{-1}[3,3].
\end{align}
Consequently, our analysis will focus on the corresponding transmission probabilities \( T_{-1} \), \( T_{0} \), and \( T_{1} \). By limiting our study to these three modes, we simplify the problem while still capturing the essential transmission behavior of the system. This approach allows us to derive meaningful results without having to deal with the full complexity of all possible transmission modes.

\section{Group delay time}\label{GDT}

The stationary state tunneling solution holds for all positions and times but does not directly reveal the dynamics of the tunneling process. To better understand the time-dependent behavior, a spatially localized wave packet can be created by combining multiple stationary states, each with different energy levels \cite{Winful2006}. This method allows a detailed study of how the wave packet evolves as it interacts with the potential barrier \cite{Ban2015}. It captures both the temporal and spatial aspects of tunneling that are not apparent in the stationary solution. It allows one to analyze the motion, propagation, and interaction of the wave packet with the barrier, providing a clearer understanding of tunneling dynamics.
Then, following the methodology outlined in \cite{Winful2006,Winful2003}, we construct a spatially localized wave packet using a Gaussian distribution $f(E - E_0)$ centered at energy $E_0 $. When this wave packet interacts with the barrier, it splits into two distinct wave packets: a reflected wave packet and a transmitted wave packet. These are expressed as
\begin{align}
			\Psi_{t}&=\int f(E-E_0)\ t(E)\ e^{\imath k_x x+\imath k_y y}\ e^{-\imath E t/\hbar}\ dE  \\
		\Psi_{r}&=\int f(E-E_0)\ r(E)\ e^{-\imath k_x x +\imath k_y y}\ e^{-\imath E t/\hbar}\ dE 
	\label{wavepacketrt}
\end{align}
where the transmission and reflection coefficients can be generally expressed as 
\begin{align}
&	t(E)=|t(E)|e^{\imath\phi_t(E)}\\
	&
	r(E)=|r(E)|e^{\imath\phi_r(E)}.
\end{align}

For a sufficiently narrow distribution function, the transmission and reflection coefficients can be treated as approximately constant over the integration range: \( |t(E)| \approx |t| \) and \( |r(E)| \approx |r| \). This implies that the wave packet retains its shape during propagation and is not significantly reshaped. In addition, its group velocity (outside the barrier) remains constant along the \( x \)-axis
\begin{equation}
	v_g=\frac{v_F \cos\phi \sqrt{(E- \lambda_{ex}^-)^2-(\lambda_{ex}^+ + \Delta)^2 }}{E - \lambda_{ex}^-}.
	\label{groupvhex}
\end{equation}
This simplification allows a clearer analysis of the behavior of the wave packet, since the constant coefficients ensure that the structure and motion of the wave packet are preserved, making it easier to study its interaction with the barrier and its propagation dynamics. Thus, dividing the barrier width by the group velocity gives the time required to travel the free space distance equal to the barrier thickness, often referred to as the equal time $t_0$.

Using the stationary phase method and assuming that the peak of the incident wave packet was initially (at time \( t = 0 \)) located at \( x = 0 \), we can determine the time required for the peak of the transmitted wave packet to emerge at the end of the barrier. This is 
\begin{equation}
	\hbar\frac{\partial}{\partial E}(\phi_t(E)+k_x D)=\tau_{gt}
	\label{groupdelay}
\end{equation}
which defines the group delay or phase time of the transmitted wave packet. A similar expression can be derived for the reflected wave packet
\begin{align}
	\hbar \frac{\partial}{\partial E} \phi_r(E)=\tau_{gr}
\end{align}
describing the time it takes for its peak to appear after reflection.
For the symmetric potentials around zero, i.e. \( V(x) = V(-x) \), both expressions give the same result, i.e., \( \tau_{gt} = \tau_{gr} \). For asymmetric potentials, it is useful to define the bidirectional group delay, such as
\begin{equation}
	\tau_g=|t|^2\tau_{gt}+|r|^2\tau_{gr}.
	\label{bidirgroupdelay}
\end{equation}	
The group delay should not be interpreted as the time required for the incident wave packet to cross the barrier. Once the wave packet begins to interact with the potential, it ceases to exist in its original form. In addition, locating the peak of the wave packet is not straightforward because the wave packet has a finite spatial extent. This means that some parts of the wave packet will reflect earlier than others, resulting in interference effects. Strictly speaking, the group delay \( \tau_{gt} \) represents the time required for the extrapolated peak of the transmitted wave packet to appear at \( x = D \), assuming that the extrapolated peak of the incident wave packet was at \( x = 0 \) at the time \( t = 0 \).



Under specific conditions, the group delay remains constant for sufficiently wide barriers. If the group delay is misinterpreted as the time required to travel a distance equal to the barrier width, a speed can be associated with the process of quantum tunneling. However, this speed is not a true measure of the motion of the wave packet through the barrier, but rather a conceptual construct derived from the group delay. It is important to note that this interpretation does not accurately reflect the physical dynamics of tunneling, as the wave packet does not propagate through the barrier in the classical sense. Instead, tunneling is a quantum phenomenon governed by wave interference and probability amplitudes that cannot be fully described by such a simplified notion of speed. This is 
\begin{equation}
	v=\frac{D}{\tau_g}.
	\label{hartmanvg}
\end{equation}
Since \( \tau_g \) approaches a constant value, this implies that \( v \) increases with barrier thickness. As there is no upper limit to the barrier width, \( v \) can reach superluminal velocities. This phenomenon is commonly referred to as the Hartman effect.


\section{Numerical result}\label{NR}

We present our results numerically to gain a deeper understanding of the factors that influence the transmission and group delay across potential barriers under laser irradiation. In particular, we study the effects of the laser field and the interactions between graphene and ferromagnetic substrates such as nickel (Ni) and cobalt (Co). In addition, we investigate the effect of the number of BN layers on the tunneling properties. By adjusting the Fermi velocity and considering the gap induced by the substrates, as shown in the table below \cite{Zollner2016}, we provide a detailed analysis by highlighting their importance in determining the transmission and group delay time. As a result, we argue that the laser field and substrate interactions are particularly crucial in shaping these two quantities, providing valuable insights into the quantum transport mechanisms in graphene-based systems.

\begin{center}\label{table1}
	\begin{tabular}{|c|c|c|c|c|}
	\hline
	Substrate& $\lambda_{ex}^A$ [meV] & $\lambda_{ex}^B$ [meV]& $\Delta$ [meV]& $v_F$ [km/s]\\
	\hline
	G/BN(1)/Ni & -1.4 & 7.78 & 22.86 &810\\
	\hline
	G/BN(2)/Ni  & 0.068 & -3.38 & 42.04 & 824\\
	\hline
	G/BN(3)/Ni  & -0.005 & 0.017 & 40.57 & 826\\
	\hline	\hline
	G/BN(1)/Co  & -3.14 & 8.59 & 19.25 &812\\
	\hline
	G/BN(2)/Co  & 0.097 & -9.81 & 36.44 & 820\\
	\hline
	G/BN(3)/Co  & -0.005 & 0.018 & 38.96 & 821\\
	\hline
\end{tabular}
\end{center}

\begin{figure}[ht]
	\centering
	\subfloat[]{\centering\includegraphics[scale=0.36]{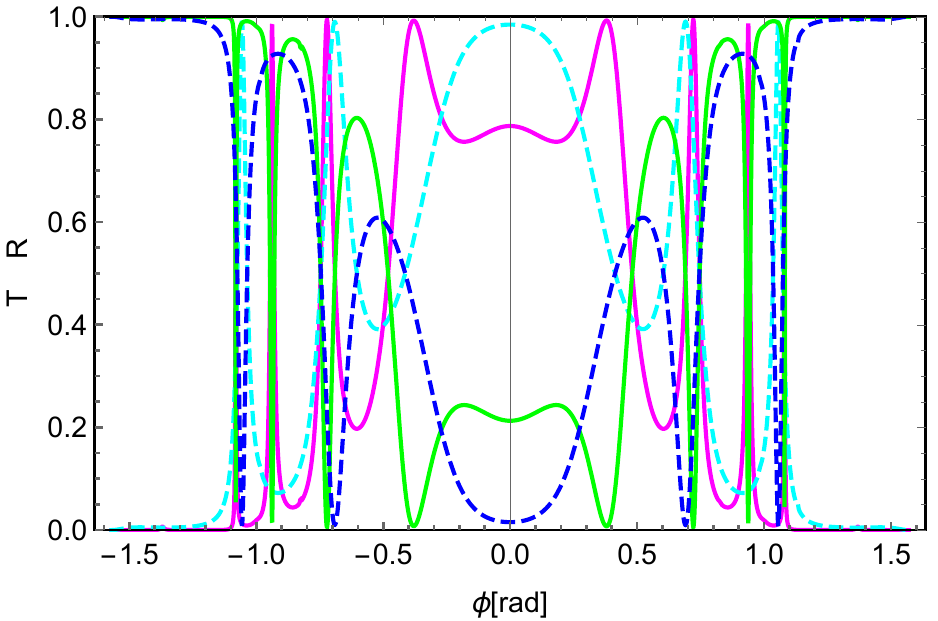}\label{fig2a}}
	\subfloat[]{\centering\includegraphics[scale=0.36]{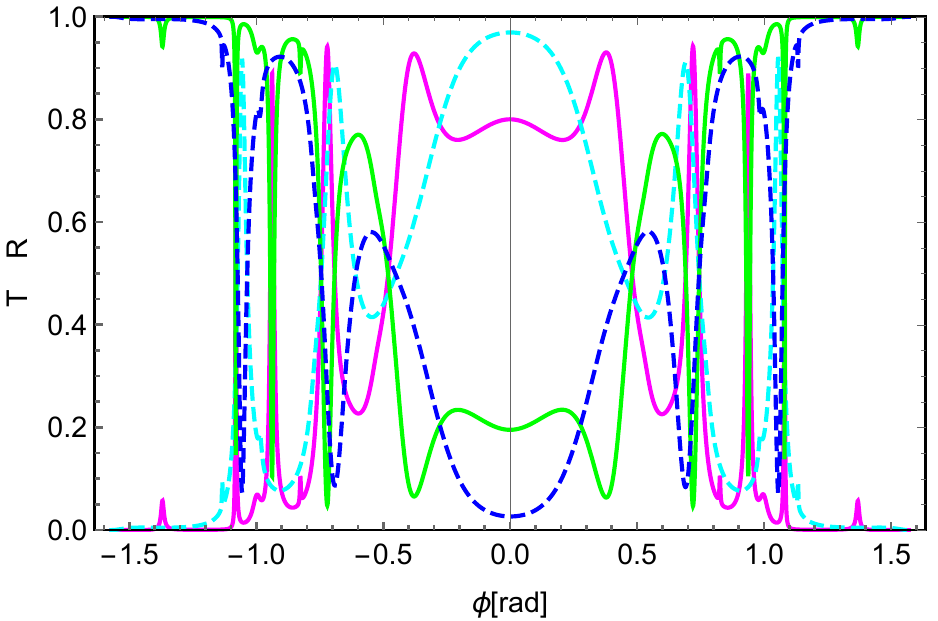}\label{fig2b}} 
	\subfloat[]{\centering\includegraphics[scale=0.36]{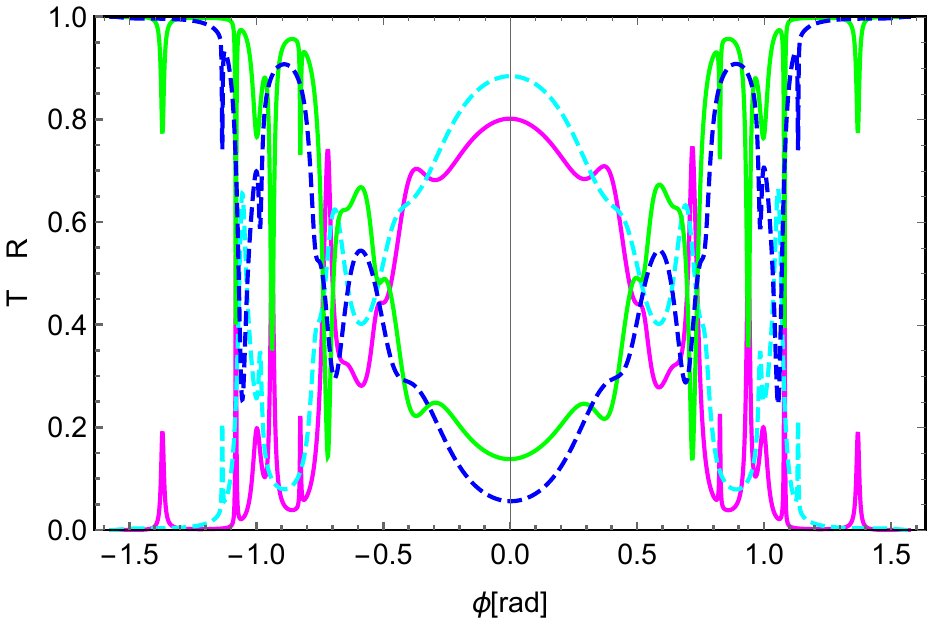}\label{fig2c}}  
	\caption{The transmission $T$ (magenta, cyan lines) and reflection $R$ (green, blue lines)  in the G/BN(1)/Ni system as a function of the incident angle $\phi$ for $\omega=12.325 \ 10^{12}$ Hz, $D=80$ nm, $E=100$ meV, and $V=200$ meV. Spin-up (solid line) and spin-down (dashed line). (a): $F=0.2{e5}$ V/m, (b): $F=0.6{e5}$ V/m, and (c): $F=1.2{e5}$ V/m.}\label{fig2}
\end{figure}

Fig. \ref{fig2} shows the transmission $T$ and reflection $R$ for the G/BN(1)/Ni system as a function of incident angle $\phi$, considering both spin-up and spin-down orientations, under the parameter configurations (\( V = 200 \) meV,\( E = 100 \) meV, \( \omega = 12.325 \ 10^{12} \) Hz,  \( D = 80 \) nm). 
{Note that the angle \(\phi\) is related to the wave vector components by the relation \(\phi = \arctan\frac{k_y}{k_x}\), where \(k_x = k_l\) as given in \eqref{eprox1}.}
 Our primary objective is to study the effect of laser irradiation on transmission through the barrier. In fact, for low value \( F = 0.2 {e5} \) V/m in Fig. \ref{fig2a}, 
we see that the transmission is zero for incident angles greater than one, indicating total reflection of the fermions. Tunneling occurs only at certain angles, and a distinct phase shift between spin-up and spin-down transmissions appears. For \( F = 0.6 {e5 } \)V/m in Fig. \ref{fig2b}, we observe that the transmission decreases and the Klein tunneling disappears. For even higher value (\( F = 1.2 {e5} \) V/m) in Fig. \ref{fig2c}, the transmission decreases further and does not exceed $80\%$. The Klein tunneling is completely suppressed, and the phase shift between spin-up and spin-down transmissions is reduced. Compared to the results obtained in \cite{Tepper2021}, we note that our results show that laser irradiation reduces the transmission and completely suppresses the Klein tunneling. 
{More precisely, we observe that the laser irradiation increases the confinement of the fermions in the potential barrier region, resulting in suppressed transmission (Fig. 2). This happened because the laser field modifies the energy spectrum of the fermions by creating additional effective barriers or wells in the system. This changes the wave vector of the fermions within the barrier region, making it more difficult for them to propagate through. As a result, the fermions experience stronger confinement, and their probability of tunneling decreases, suppressing transmission. This effect is consistent with the laser-induced modulation of electronic properties in graphene, as discussed in \cite{Oka2009, Kibis2010}.}

\begin{figure}[ht]
	\centering
	\subfloat[]{\centering\includegraphics[scale=0.36]{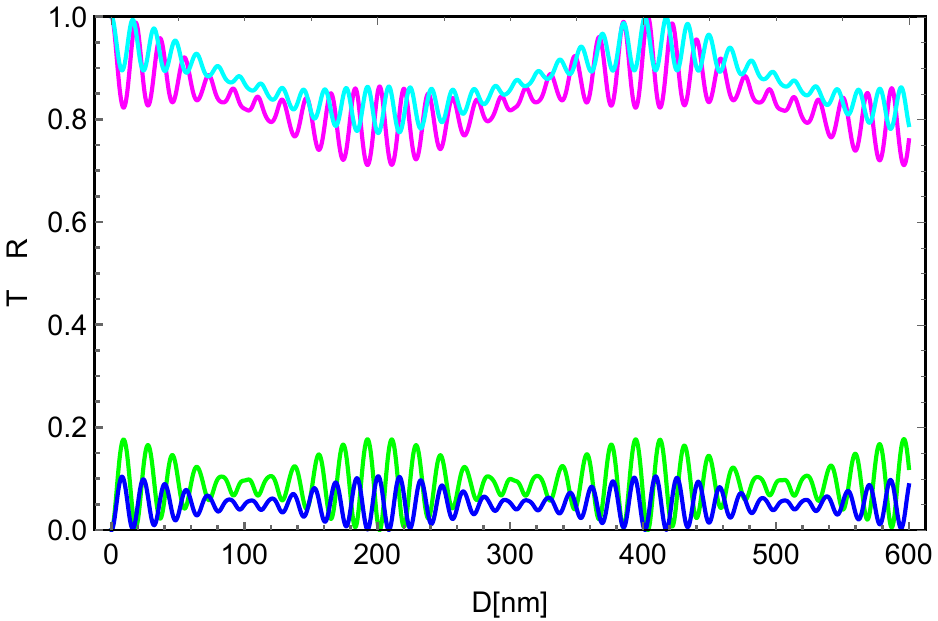}\label{fig22a}}
	\subfloat[]{\centering\includegraphics[scale=0.36]{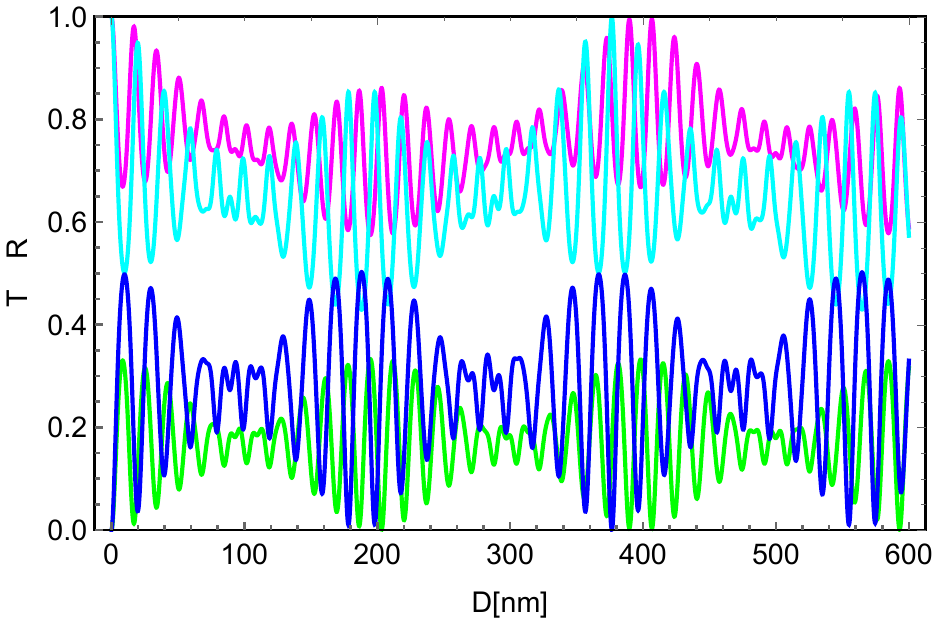}\label{fig22b}} 
	\subfloat[]{\centering\includegraphics[scale=0.36]{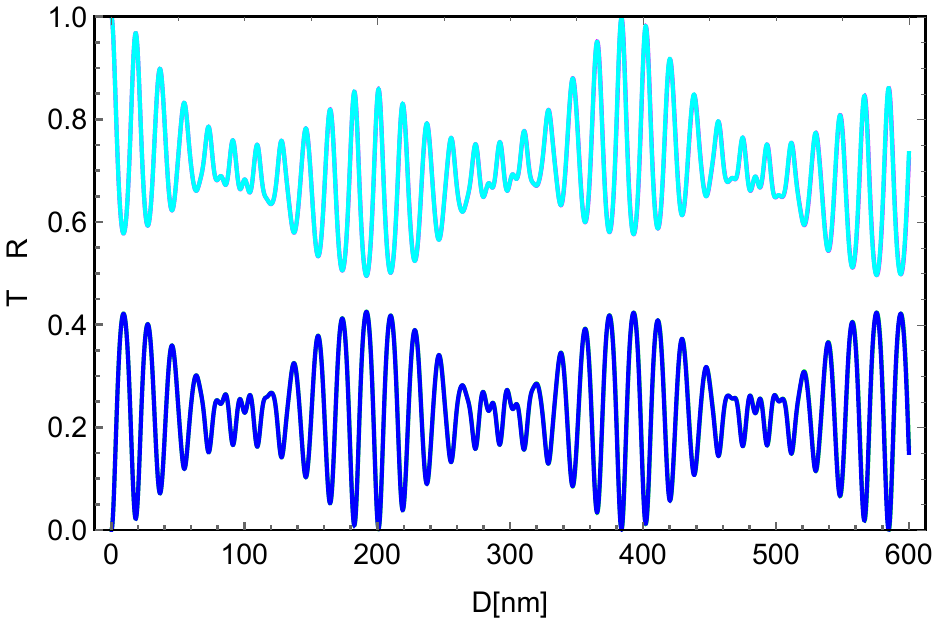}\label{fig22c}}  
	\caption{The transmission $T$ (magenta, cyan lines) and reflection $R$ (green, blue lines)  in the G/BN($j$)/Co system as a function of the barrier width $D$ at normal incidence for $F=1.2{e5}$ V/m, $\omega=12.325 \ 10^{12}$ Hz, $E=100$ meV, and $V=300$ meV.  Spin-up (magenta, blue lines) and spin-down (cyan, green lines). (a): $j=1$, (b): $j=2$, and (c): $j=3$.}\label{fig22}
\end{figure}

Fig. \ref{fig22} shows the transmission $T$ as a function of the barrier width \( D \) for graphene on a nickel substrate separated by boron nitride (BN) layers under a strong laser field. The goal is to study how the number of BN layers and the barrier width affect the transmission. In Fig. \ref{fig22a}, with one BN layer ($j=1$), the transmission oscillates around unity. Total transmission occurs near \( D = 400 \) nm, indicating resonant tunneling. In Fig. \ref{fig22b}, the presence of two BN layers (\( j = 2 \)) results in beat oscillations in the transmission. In this case, the phase shift between spin-up and spin-down transmissions becomes smaller. This shows that the additional BN layer introduces interference effects. With three BN layers (\( j = 3 \)) in Fig. \ref{fig22c}, the phase shift between spin-up and spin-down transmissions disappears. 
The oscillations become stronger, showing that more BN layers increase more changes in the system properties. These results show that the transmission depends on the number of BN layers. Beat oscillations are always present, proving that interference plays a key role. The results also show how BN layers affect spin-dependent behavior and transmission. This is useful for the design of graphene-based devices where control of spin and transmission is important.
{Note that the proximity exchange effect, like the interlayer exchange coupling in magnetic multilayers, is greatly reduced with three BN layers \cite{Tepper2021}. This reduction in spin-dependent effects explains why the phase shift in Fig. 3c disappears.}

\begin{figure}[ht]
	\centering
	\subfloat[]{\centering\includegraphics[scale=0.36]{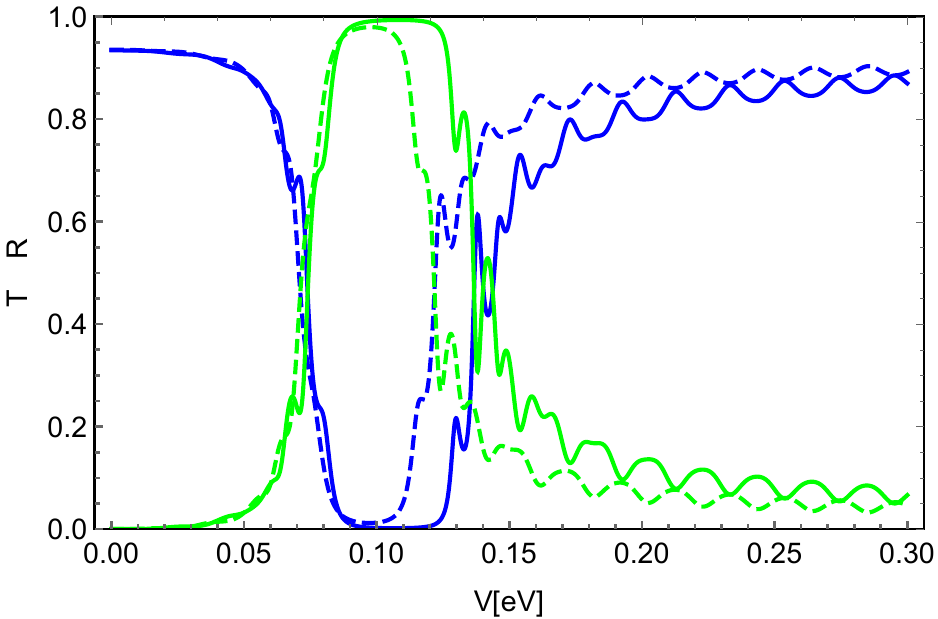}\label{fig3a}}
	\subfloat[]{\centering\includegraphics[scale=0.36]{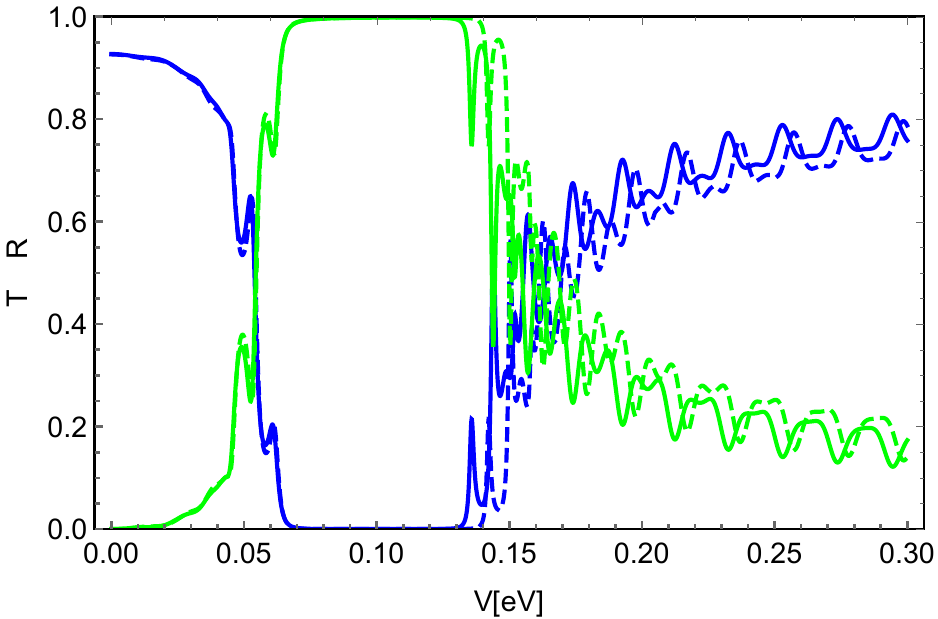}\label{fig3b}} 
	\subfloat[]{\centering\includegraphics[scale=0.36]{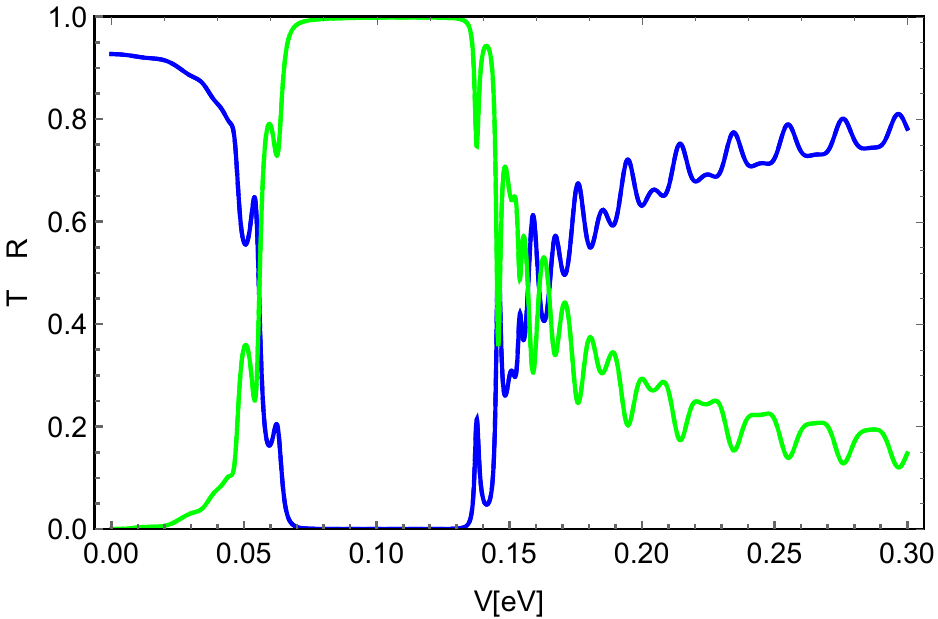}\label{fig3c}} \\
	\caption{The transmission $T$ (blue) and reflection $R$ (green)  in the G/BN($j$)/Ni $(j=1,2,3)$ system as a function of the barrier height $V$ at normal incidence for  $F=1.23{e5}$ V/m, $\omega=12.325 \ 10^{12}$ Hz, $D=80$ nm, and $E=100$ meV. Spin-up (solid line) and spin-down (dashed line). (a): G/BN(1)/Ni,  (b): G/BN(2)/Ni, and (c): G/BN(3)/Ni.}\label{fig3} 
	\end{figure}	
	
Fig. \ref{fig3} shows the transmission $T$ and reflection $R$ in the G/BN(j)/Ni (j=1,2,3) system as a function of the barrier height $V$ at normal incidence for \( F = 1.23 {e5}\) V/m, \( \omega = 12.325 \ 10^{12} \) Hz, \( D = 80 \) nm, and \( E = 100 \) meV. For a single layer of BN in Fig. \ref{fig3a},  we observe that the transmission is zero near \( V = 100 \) meV. Outside this region, transmission is not total, even when the incident energy exceeds the barrier height, due to the effect of laser irradiation. 
For two BN layers in Fig. \ref{fig3b}, the width of the zero transmission region increases. {It is evident that the transmission drops to zero when the potential \(V\) is in the range (\(\beta \lambda_B + E + l\hbar\omega - \Delta\)) and (\(\beta \lambda_B + E + l\hbar\omega + \Delta\)). This indicates a suppression of the transmission due to the specific energy conditions imposed by the system parameters. We observe that as the barrier height \( V \) increases, the phase shift between the spin-up and spin-down transmissions decreases, and the transmission increases. This occurs because higher barrier heights can create resonance conditions where electronic waves interact constructively, enhancing transmission by facilitating the passage of electrons through the barrier}.
For three layers of BN in Fig. \ref{fig3c}, the transmission varies in the same way as before, but with the phase shift disappearing. These results show that laser irradiation can reduce the transmission due to the degeneracy of the energy levels of the fermions, allowing them to be trapped between these levels. As a result, the Klein tunneling disappears, unlike in the case of a simple barrier \cite{Tepper2021}, where such tunneling persists even when the energy of the fermions is less than the barrier height.
When \( V \) is close to \( E \), the transmission drops to zero, resulting in total reflection of the fermions. The width of the zero transmission region increases as more BN layers are added. In addition, the transmission depends on the spin, with a clear difference between spin-up and spin-down states. This spin dependence weakens with one BN layer and disappears with three BN layers, leading to spin-independent transmission. Transmission also changes with barrier height, with resonance peaks becoming more pronounced as \( V \) increases. This behavior differs from pristine graphene, where transmission is largely unaffected by barrier height \cite{Klein1929}. These results highlight the unique properties of our structure compared to single barriers \cite{Tepper2021} and double barriers \cite{Jellal2024}.

\begin{figure}[ht]
	\centering
	\subfloat[]{\centering\includegraphics[scale=0.35]{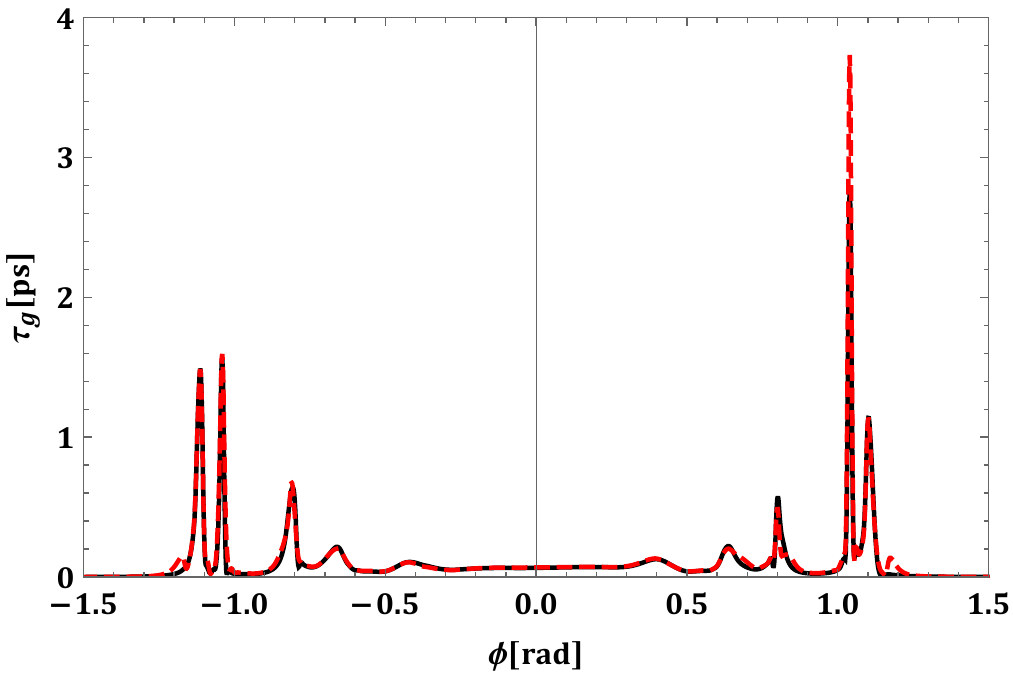}\label{fig4a}}
	\subfloat[]{\centering\includegraphics[scale=0.35]{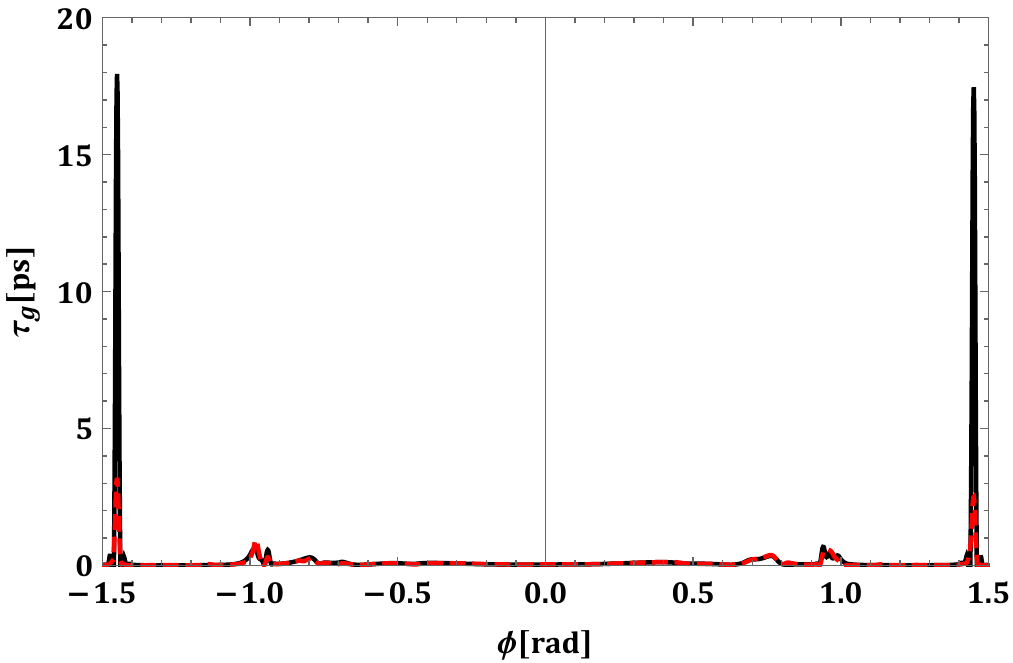}\label{fig4b}} 
	\subfloat[]{\centering\includegraphics[scale=0.35]{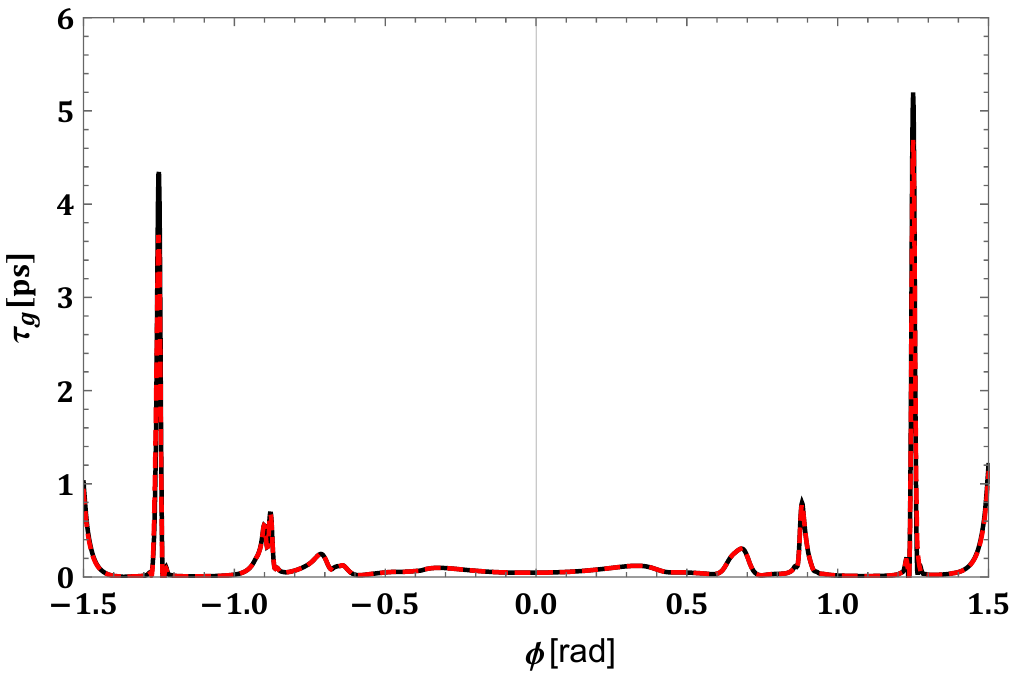}\label{fig4c}} \\
	\subfloat[]{\centering\includegraphics[scale=0.35]{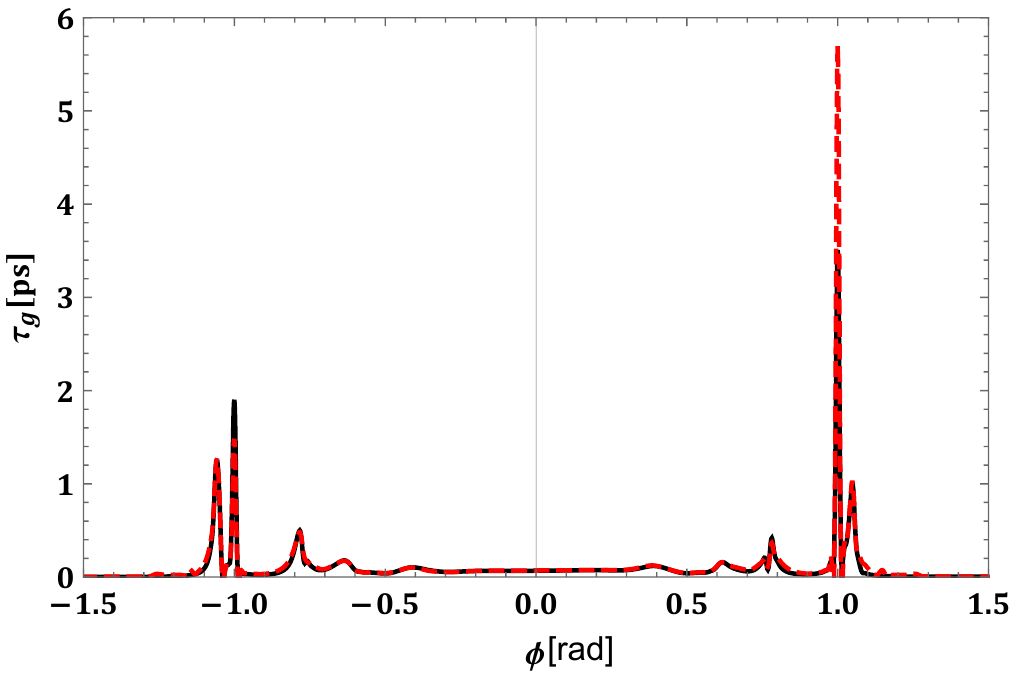}\label{fig4d}}
	\subfloat[]{\centering\includegraphics[scale=0.35]{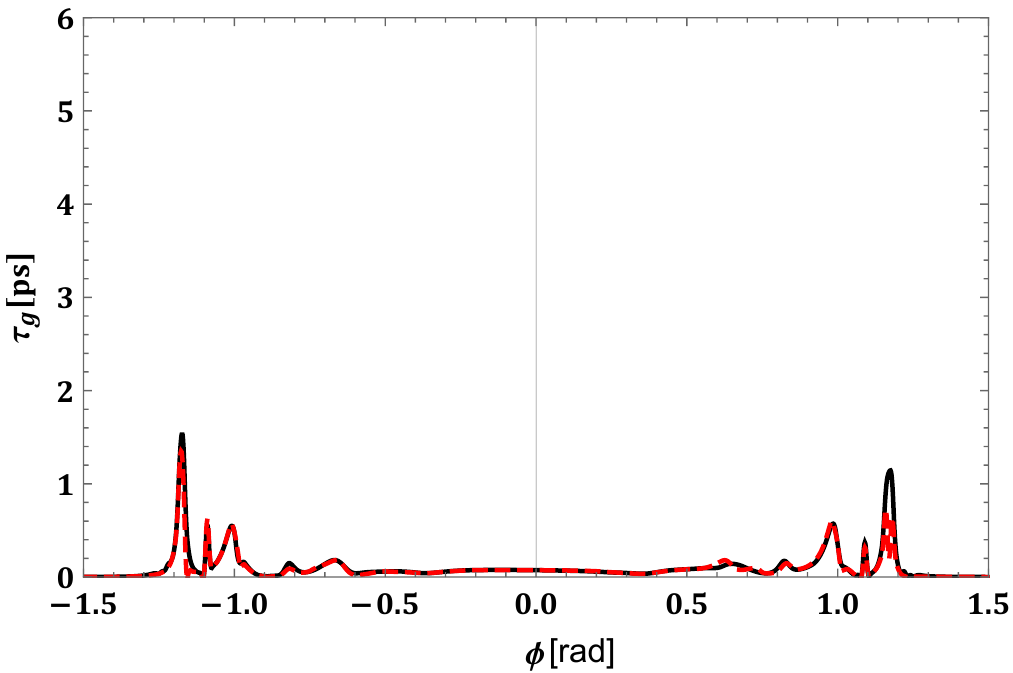}\label{fig4e}} 
	\subfloat[]{\centering\includegraphics[scale=0.35]{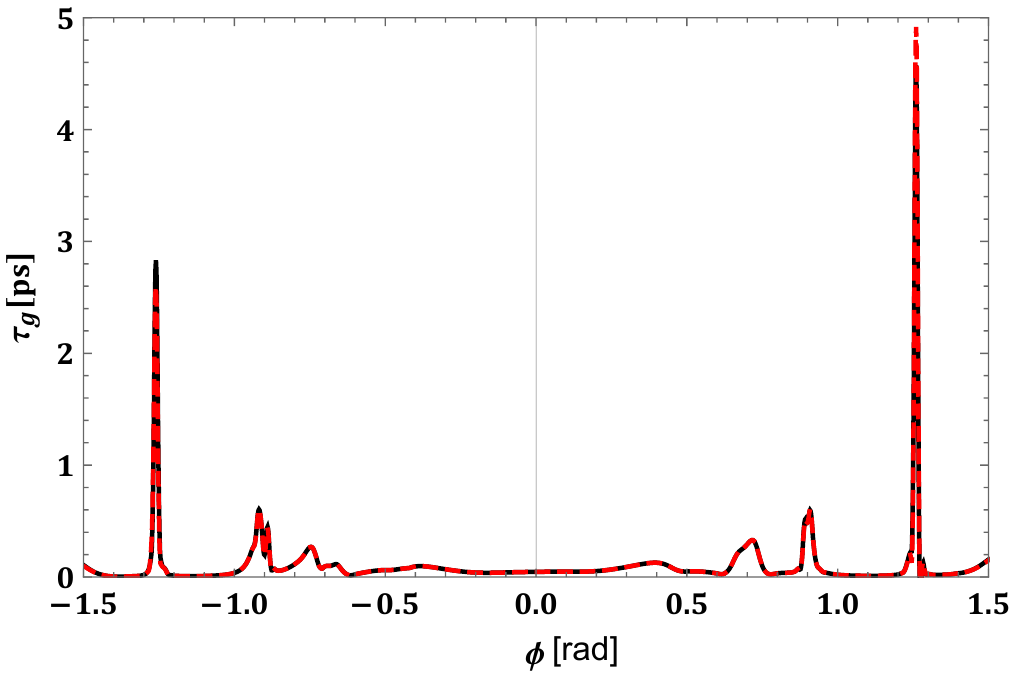}\label{fig4f}} 
	\caption{The group delay time $\tau_g[s]$ as a function of incident angle $\phi$ for  $\omega=12.325 \ 10^{12}$ Hz, $D=60$ nm, $E=100$ meV, {$F=1.2e5$} V/m and $V=300$ meV. Spin-up (black line) and spin-down (red dashed line). First line G/BN(j)/Ni system and second line for G/BN($j$)/Co system $(j=1,2,3)$.}\label{fig4}
\end{figure}

Fig. \ref{fig4} shows the group delay time $\tau_g[s]$ as a function of the incident angle $\phi$ for the parameters $V=200$ meV, $E=100$ meV, $d=60$ nm, $F=1.23{e5}$ V/m, and $\omega=1.2\ 10^{12}$ Hz. The first line corresponds to the nickel substrate and the second to the cobalt substrate. Fig. \ref{fig4a} shows a single layer of BN, where we observe that the group delay time is zero for most angles, with a maximum value close to {$4$ps}. In addition, new peaks appear compared to the case without laser irradiation \cite{Tepper2021, Jellal2024}. The spin-up and spin-down transmissions are nearly identical, with only a small difference in amplitude. This indicates the absence of proximity exchange coupling after laser irradiation. For two layers of BN, Fig. \ref{fig4b} shows that the group delay time reaches values close to $15$ps, but it is zero for most angles.
For three BN layers in Fig.~\ref{fig4c}, the group delay time does not exceed $5$ps. For the case of the cobalt substrate, the same observation holds—the group delay time is zero for most incident angles, as shown in Figs. (\ref{fig4d},\ref{fig4e},\ref{fig4f}).
For three BN layers in Fig.~\ref{fig4c}, the group delay time does not exceed $5$ps. For the case of the cobalt substrate, the same observation holds—the group delay time is zero for most incident angles, as shown in Figs. (\ref{fig4d},\ref{fig4e},\ref{fig4f}).
As we have discussed previously \cite{Elaitouni2022, ELAITOUNI2024, Elaitouni2023A, doublelaser, Elaitouni2023}, laser irradiation enhances the interaction between fermions and the field. This creates extra subbands in the energy spectrum. Thus, fermions become trapped between these levels, which is an effect known as the Stark effect \cite{Stark}. This results in greater fermion confinement, reducing transmission and potentially delaying fermions within the barrier. Laser irradiation also breaks the symmetry observed in the absence of a laser field, where the peaks are symmetric \cite{Tepper2021, Jellal2024}, and further reduces spin coupling.

\begin{figure}[ht]
	\centering
	\subfloat[]{\centering\includegraphics[scale=0.31]{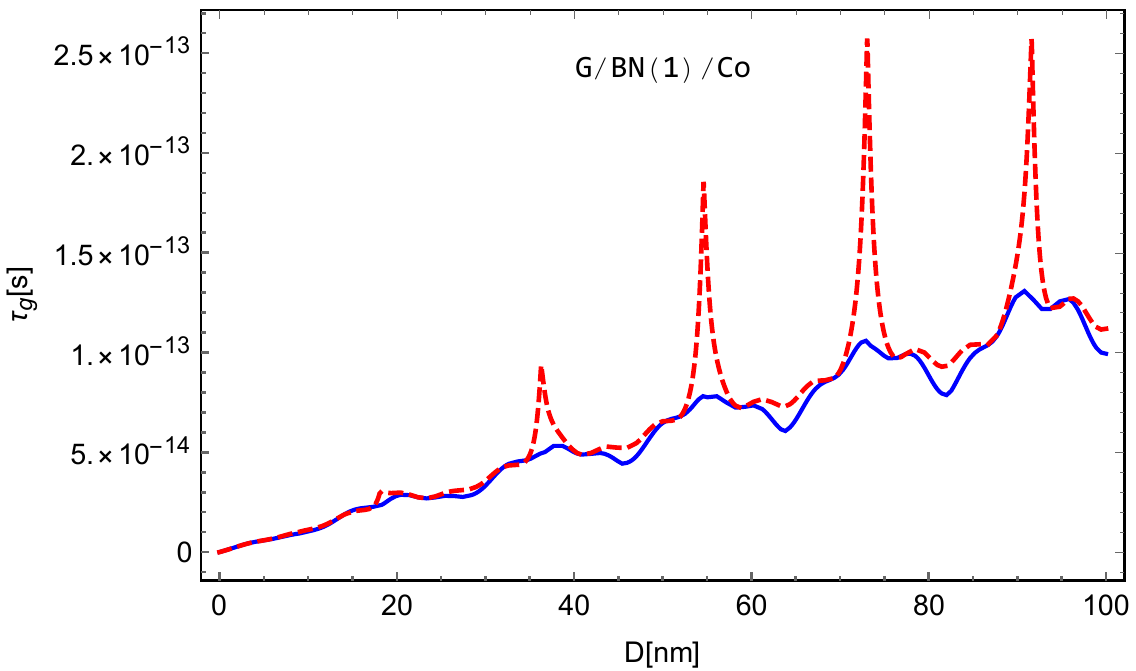}\label{fig5a}}
	\subfloat[]{\centering\includegraphics[scale=0.31]{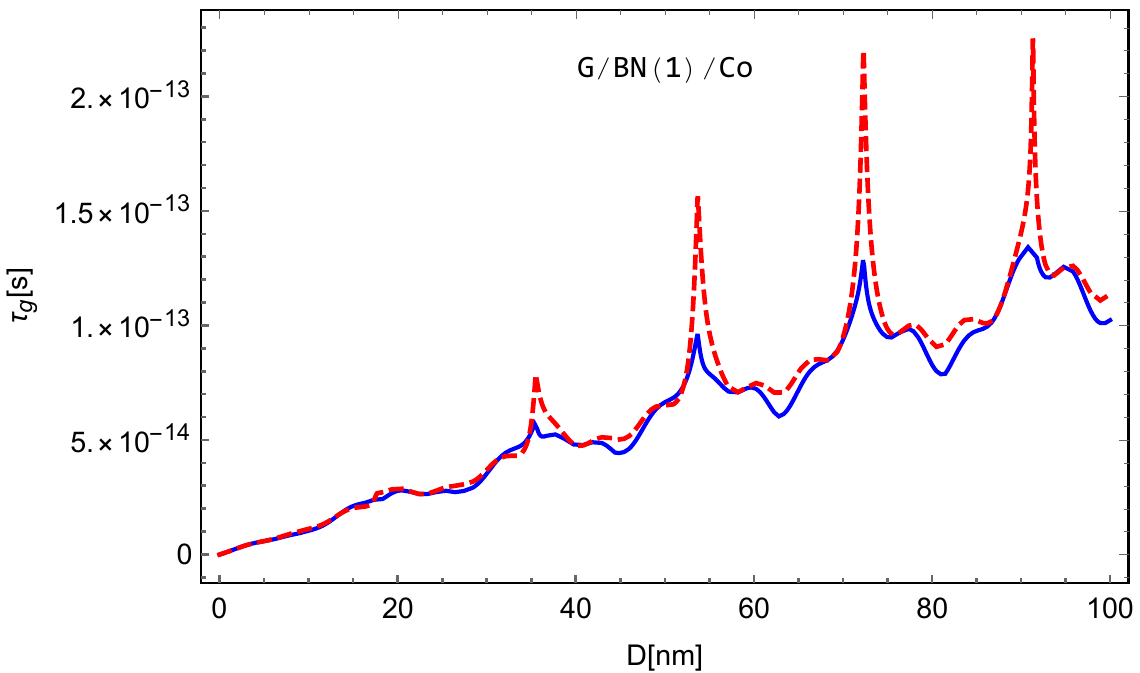}\label{fig5b}} 
		\subfloat[]{\centering\includegraphics[scale=0.31]{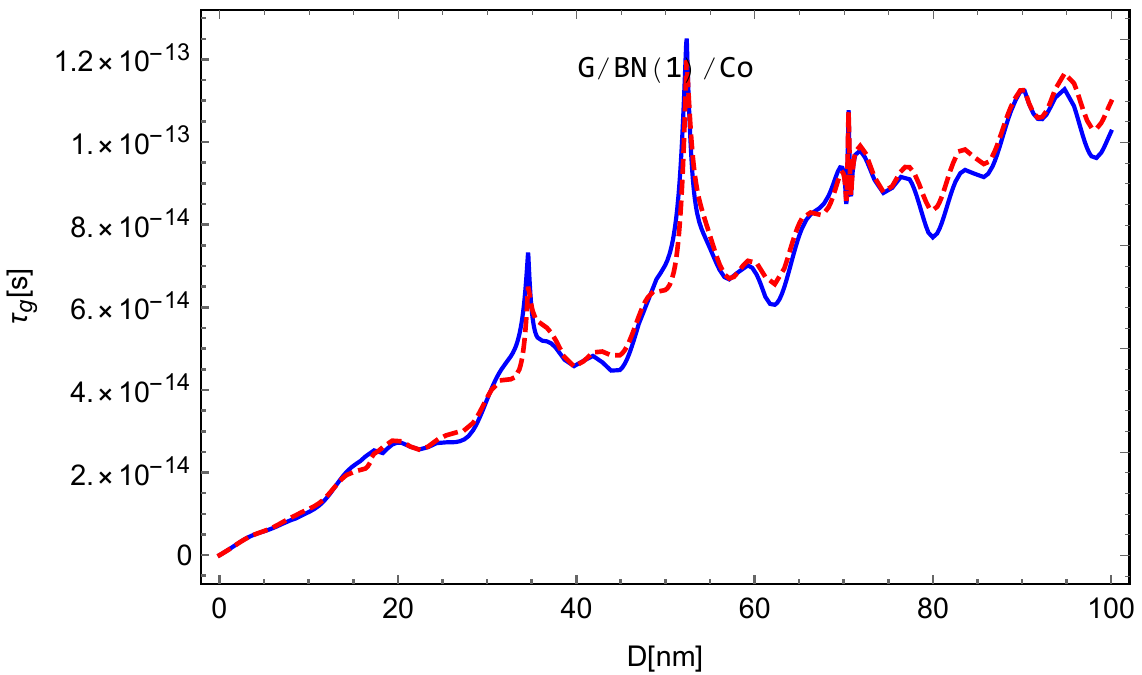}\label{fig5c}} 
	\caption{The group delay time $\tau_g[s]$ in G/BN(1)/Co system  as a function of the barrier width $D$  at normal incidence for $\omega=12.325 \ 10^{12}$ Hz, $E=100$ meV, and $V=200$ meV. Spin-up (blue line) and spin-down (red dashed line). (a): $F=0.2{e5}$ V/m, (b): $F=0.6{e5}$ V/m. (c): $F=1.2{e5}$ V/m.}\label{fig5}
\end{figure}

Fig. \ref{fig5} shows the group delay time \(\tau_g[s]\) as a function of the barrier width \(D\) for the nickel substrate with a single BN layer at normal incidence under different laser intensities \(F\). For \(F = 0.2 {e5}\) V/m, Fig. \ref{fig5a} shows that the group delay time increases with increasing barrier width. This is expected because the delay time is proportional to the traveled distance. 
The phase shift between spin-up and spin-down transmissions is clearly visible, as shown by the blue and red curves.
For \( F = 0.6 {e5}\) V/m in Fig. \ref{fig5b}, the group delay time follows a similar pattern to the previous case. However, there is a noticeable decrease in amplitude and coupling dephasing.
When a stronger field is applied, as shown in Fig. \ref{fig5c}, the effect of laser irradiation on spin coupling and group delay time becomes more pronounced. This confirms that the group delay time depends on the barrier width, indicating the absence of the Hartman effect \cite{Hartman1962}. Increasing the laser intensity reduces both the spin coupling effect and the group delay time for normal incidence.

\begin{figure}[ht]
	\centering
	\subfloat[]{\centering\includegraphics[scale=0.36]{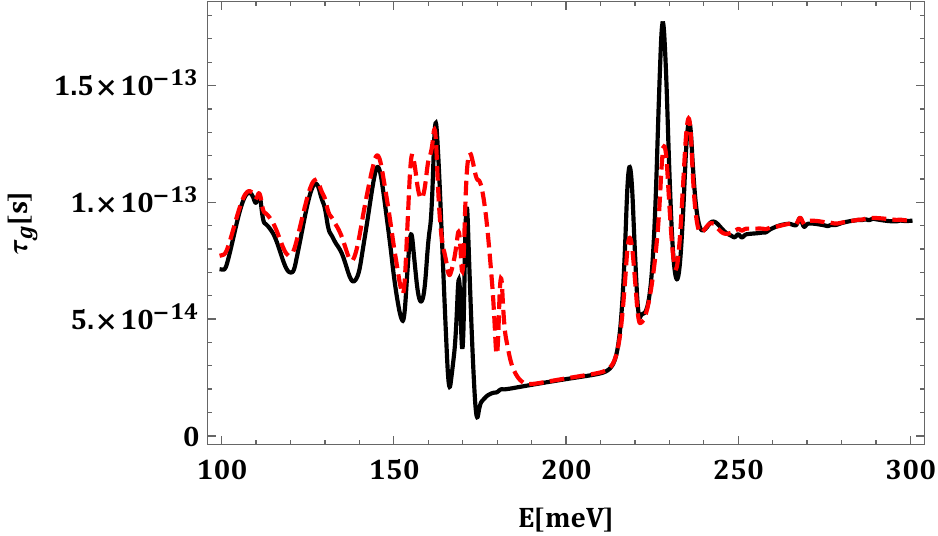}\label{fig6a}}
	\subfloat[]{\centering\includegraphics[scale=0.36]{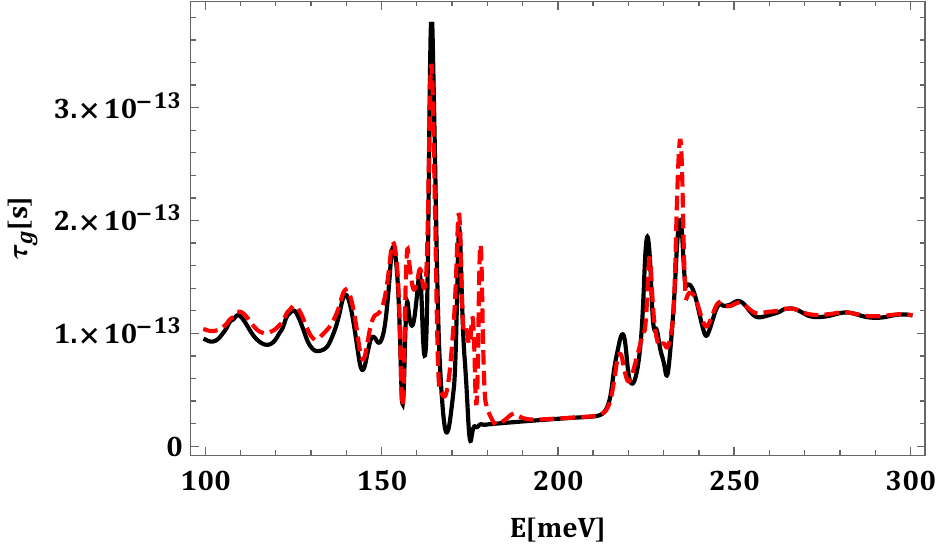}\label{fig6b}} 
	\subfloat[]{\centering\includegraphics[scale=0.36]{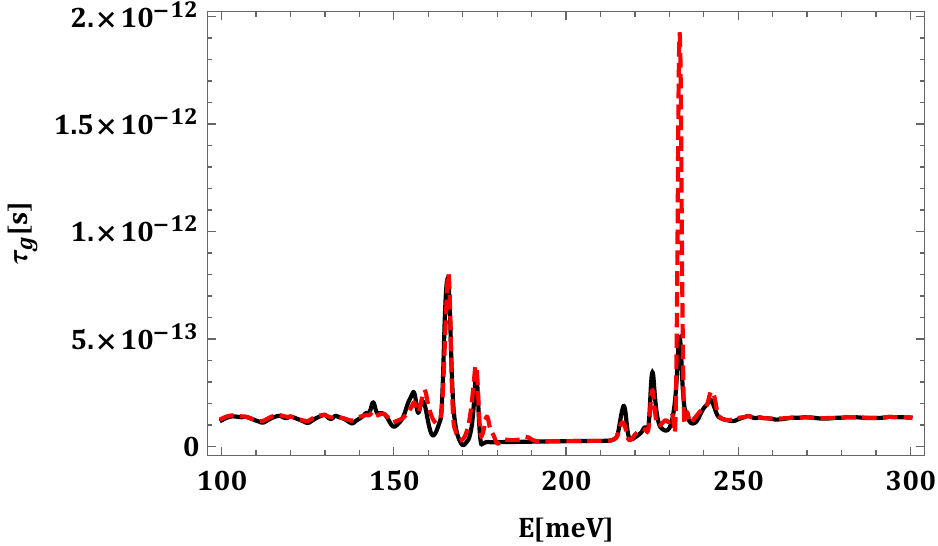}\label{fig6c}} 
	\caption{The group delay time $\tau_g[s]$ in the G/BN(1)/Ni system 
		as a function of incident energy $E$
		at
	 normal incidence for $\omega=12.325 \ 10^{12}$ Hz, $F=1.2{e5}$ v/m and $V=200$ meV. Spin-up (black line) and spin-down (red dashed line). (a): $D=80$ nm, (b): $D=100$ nm, and (c): $D=120$ nm.}\label{fig6}
\end{figure}

Fig. \ref{fig6} shows the group delay time $\tau_g[s]$ as a function of incident energy $E$ at normal incidence for graphene deposited on a nickel substrate with a single BN interlayer, under  the laser field of {$F=1.2e5$ V/m} and  $\omega=1.3\ 10^{12}$ Hz ($\alpha$ close to $1$). For the barrier width  $D=80$ nm in Fig. \ref{fig6a}, we  observe that the group delay time oscillates for energies less than $V$. The oscillations then disappear for a short time and then reappear. The maximum time does not exceed the value $0.15$ ps for $E < V$, but exceeds it for $E > V$. After $250$ meV, the group delay time stabilizes around the value $0.1$ ps. For $D=100$ nm in Fig. \ref{fig6b}, the group delay time follows a similar pattern, reaching its maximum for $E < V$ and exceeding the time value $3$ ps. Again, the time also stabilizes around the value $0.1$ ps after $E > 250$ meV.  For $D=120$ nm in Fig. \ref{fig6c}, the group delay  delay still varies in the same way with a slight increase. For an incident energy  $E=230$ meV, the group delay time reaches a maximum value close to $2$ ps for spin down. We note that  increasing the barrier width increases the group delay time for normal incidence, implying the absence of the Hartman effect. Laser irradiation causes more delay within the barrier due to the interaction between fermions and the field, with stabilization after a certain energy.

\section{Conclution}\label{Con}

We have studied the confinement of fermions in graphene on ferromagnetic substrates such as nickel or cobalt. A static potential barrier with height \( V \) and width \( D \) was used, and the system was irradiated with a monochromatic laser field with linear polarization. The Floquet approximation was used to determine the wave function because the laser field oscillates with time. By ensuring the continuity of the wave function at the barrier boundaries, we derived a system of four equations with an infinite number of modes. To simplify the analysis, we used matrix formalism and current densities to calculate transmission and reflection probabilities. Our study focused on the first two bands: the central band with energy \( E \) and the first sideband with energy \( E \pm \hbar\omega \). The Klein tunneling effect was observed only in weaker laser fields and disappeared with increasing laser intensity.

The laser irradiation intensity inversely affects transmission due to the interaction between fermions and the field. As the laser intensity increases, the transmission probability decreases because fermions become trapped between the energy subbands created by the field. Laser irradiation also significantly reduces the spin coupling effect, leading to more uniform transmission for both spin-up and spin-down states. This effect is further enhanced by adding boron nitride (BN) layers, which not only diminish the spin dependence of transmission but also widen the range where transmission is canceled. Additionally, laser irradiation decreases the group delay time for normal incidence while increasing it for other angles. The group delay time depends on the barrier width, unlike the Hartman effect, where the delay time remains constant regardless of barrier thickness. In our setup, the Hartman effect is absent. The inclusion of BN layers further reduces the spin effect on transmission and extends the range of transmission cancellation. This graphene-based structure allows for effective confinement of fermions and precise control over their traversal time through the barrier.

{Our theoretical results can help to understand and control electron transport in graphene-based devices, which are important for nanoelectronics and spintronics. For example, the interaction between the potential barrier and the laser field can be used to design tunable graphene transistors with better performance \cite{Schwierz2010}. The use of ferromagnetic substrates such as cobalt or nickel also opens the door to graphene-based spintronic devices, where spin-polarized currents can be controlled for low-power memory and logic applications \cite{Han2014}. Experimental techniques such as electrostatic gating and laser modulation are already being used in advanced research to fabricate and study graphene heterostructures \cite{Dean2010, Oka2009}. These developments demonstrate the practical value of our theoretical framework and its potential to guide future experiments to make the most of the unique properties of graphene for new technologies.}

 \section*{Acknowledgment}
P.D. and D.L. acknowledge partial financial support from FONDECYT 1231020.

\end{document}